\documentclass[12pt]{aastex631}

\usepackage{graphicx}

\definecolor{purple}{rgb}{1,0,1}

\newcommand{\cxo}{{\it Chandra}}
\newcommand{\rxte}{{\it RXTE}}

\newcommand{\nustar}{{\it NuSTAR}}

\newcommand{\swift}{{\it Swift}}

\newcommand{\nicer}{{\it NICER}}

\newcommand{\psra}{{PSR J1846$-$0258}}

\shorttitle{J1846$-$0258 2020 Outburst}
\shortauthors{Hu et al.}
\graphicspath{{./}{figures/}}

\def\be{\begin{equation}}
\def\ee{\end{equation}}

\def\lsim{\lower 2pt \hbox{$\, \buildrel {\scriptstyle <}\over
         {\scriptstyle \sim}\,$}}

\def\psr{J1846$-$0258 }
\def\eins{1E1547.0$-$5408}

\begin{document}

\title{A \nicer\ view on the 2020 Magnetar-Like Outburst of PSR J1846$-$0258}

\author{Chin-Ping Hu}
\affiliation{Department of Physics, National Changhua University of Education, Changhua 50007, Taiwan}
\author{Lucien Kuiper}
\affiliation{SRON-Netherlands Institute for Space Research, Niels Bohrweg 4, 2333 CA, Leiden, The Netherlands}
\author{Alice K. Harding}
\affiliation{Theoretical Division, Los Alamos National Laboratory, Los Alamos, NM 87545 USA}
\author{George Younes}
\affiliation{Department of Physics, The George Washington University, Washington, DC 20052, USA} 
\affiliation{Astronomy, Physics and Statistics Institute of Sciences (APSIS), The George Washington University, Washington, DC 20052, USA}
\author{Harsha Blumer}
\affiliation{Department of Physics and Astronomy, West Virginia University, Morgantown, WV 26506, USA}
\affiliation{Center for Gravitational Waves and Cosmology, West Virginia University, Chestnut Ridge Research Building, Morgantown, WV 26505, USA}
\author{Wynn C. G. Ho}
\affiliation{Department of Physics and Astronomy, Haverford College, 370 Lancaster Avenue, Haverford, PA 19041, USA}
\author{Teruaki Enoto}
\affiliation{Extreme Natural Phenomena RIKEN Hakubi Research Team, Cluster for Pioneering Research, RIKEN, 2-1 Hirosawa, Wako, Saitama 351-0198, Japan}
\author{Crist\'obal M. Espinoza}
\affiliation{Departamento de F\'isica, Universidad de Santiago de Chile (USACH), Av. Victor Jara 3493, Estaci\'on Central, Chile}
\affiliation{Center for Interdisciplinary Research in Astrophysics and Space Sciences (CIRAS), Universidad de Santiago de Chile, Santiago, Chile}
\author{Keith Gendreau}
\affiliation{X-Ray Astrophysics LAboratory, NASA Goddard Space Flight Center, Greenbelt, MD 20771}

\correspondingauthor{A. K. Harding}
\email{ahardingx@yahoo.com}

\begin{abstract}
We report on our monitoring of the strong-field magnetar-like pulsar PSR \psr with the Neutron Star Interior Composition Explorer (\nicer) and the timing and spectral evolution during its outburst in August 2020.  Phase-coherent timing solutions were maintained from March 2017 through November 2021, including a coherent solution throughout the outburst.  We detected a large spin-up glitch of magnitude $\Delta\nu/\nu = 3 \times 10^{-6}$ at the start of the outburst and observed an increase in pulsed flux that reached a factor of more than 10 times the quiescent level, a behavior similar to that of the 2006 outburst.  Our monitoring observations in June and July 2020 indicate that the flux was rising prior to the \swift\ announcement of the outburst on August 1, 2020.  We also observed several sharp rises in the pulsed flux following the outburst and the flux reached quiescent level by November 2020.  The pulse profile was observed to change shape during the outburst, returning to the pre-outburst shape by 2021.  Spectral analysis of the pulsed emission of \nicer\ data shows that the flux increases result entirely from a new black body component that gradually fades away while the power-law remains nearly constant at its quiescent level throughout the outburst.  Joint spectral analysis of \nicer\ and simultaneous \nustar\ data confirms this picture.  We discuss the interpretation of the magnetar-like outburst and origin of the transient thermal component in the context of both a pulsar-like and a magnetar-like model.

\end{abstract}



\section{Introduction} \label{sec:intro}

The dividing line between the two neutron star populations, rotation-powered pulsars (RPPs) and magnetars, has been blurring over the past ten years with the discovery of magnetar-like outbursts from the RPPs PSR \psr and J1119-6127.  Both of these pulsars possess very strong surface dipole magnetic fields that are at the upper end of what is observed for RPPs, with $5 \times 10^{13}$ G for \psr  and $4 \times 10^{13}$ G for J1119-6127.  The first observed outburst from \psr occurred in 2006 \citep{Kumar2008,Gavriil2008}, during which the pulsed flux increased by a factor of five, decaying over several months, and a transient thermal component appeared in addition to the persistent non-thermal emission, whose flux rose by 35\% \citep{Kuiper2009}.  PSR J1119$-$6127 experienced a magnetar-like outburst in 2016 \citep{Archibald2016}, but in this case a transient non-thermal power law component appeared and the quiescent purely thermal pulsed flux increased by a factor of $> 160$.  In both outbursts, large spin-up glitches occurred, with $\Delta\nu/\nu = 5.74 \times 10^{-6}$ for J1119$-$6127 and $\Delta\nu/\nu = 2 - 4.4 \times 10^{-6}$  for \psr \citep{Livingstone2010}.

PSR \psr was discovered through 0.324 s X-ray pulsations by \citet{Gotthelf2000} and with its high spin-down rate of $\dot P = 7 \times 10^{-12}\,\rm s\,s^{-1}$ has a characteristic age of $\tau = 723$ yr, making it the youngest known RPP.  It is located in the Kes 75 supernova remnant and surrounded by a compact pulsar wind nebula \citep{Helfand2003, NgSG2008}.  Its pulsed emission is still only seen in the X-ray band from $3 - 200$ keV with \rxte\, INTEGRAL,  \citep{Kuiper2015}, and recently also in soft $\gamma$-rays below 100 MeV with Fermi \citep{Kuiper2018}.  Its spectral energy distribution (SED) is very similar to that of PSR B1509$-$58, another young, energetic RPP, with an SED peak around 1 - 10 MeV and a decline above 30 MeV \citep{Kuiper2018}.  However, B1509$-$58 is a radio pulsar while \psr is radio quiet.  These two pulsars are members of a class of so-called ``MeV pulsars" whose SEDs peak at MeV energies, contrary to most other energetic RPPs peaking at GeV energies. The MeV pulsars are a group of 11 rotation-powered pulsars that exhibit strong, non-thermal hard X-ray emission, have broad single pulse profiles, and no detected gamma-ray pulsations above 1 GeV \citep{Kuiper2018,Harding2017}.  All except one are radio quiet. 

In this paper, we will describe our long-term monitoring and timing of \psr using \nicer\ (and \swift\ for the pre-2018 era) in Section \ref{sec:mon}.  As a result of this program we have obtained phase-coherent timing solutions from April 2017 through November 2021.  We will also report on the magnetar-like outburst of August 2020 and its aftermath, during which we obtained Target-of-Opportunity (ToO) observations following the \swift\ announcement \citep{Krimm2020} of a magnetar-like $\sim 0.1$ s flare/burst.  Near the start of the outburst we detected a large spin-up glitch \citep{Kuiper2020} and maintained coherent timing over the entire outburst.  Timing through the outburst and the evolution of the pulsed flux and pulse profile are described in Section \ref{sec:timing}.  The spectral evolution before and during the outburst are discussed in Section \ref{sec:spectra}.  Finally in Section \ref{sec:diss} we will discuss the implications of these results and how \psr fits in to the RPP-magnetar connection.

\section{Observations and Data Reduction}  \label{sec:mon}

\subsection{\nicer\ Monitoring}

We have been timing \psr with \nicer\ starting on May 23, 2018 (Observation id. 1033290101; MJD 58261) through a series of Guest Observer proposals, AO1 - AO4 GO.
The last \nicer\ observation used in this work was performed on November 13, 2021 (Observation id. 4607021501; MJD 59531), and so our \nicer\ monitoring covered a $\sim 3.5$ year period subjected to $\sim 3$ months (approximately lasting from end November till end February) regular episodes in which no data can be taken due to visibility constraints resulting in (large) data gaps.

The \nicer\ data were processed using HEASOFT version 6.27.2 and NICERDAS version 7a. We used the latest calibration files obtained from the standard CALDB release for \nicer\, downloaded from NASA’s High Energy Astrophysics Science Archive Research Center (HEASARC).  We start our data reduction from level 1 event files. We create good time intervals using standard filtering criteria as described in the \nicer\ Data Analysis Guide. For spectral studies, we excluded module numbers 14 and 34 to avoid potential contamination by instrumental noise in the soft energy band. 

For the timing analysis, requiring less restrictive selections as for the spectral analysis, we used predominantly the screened event files as stored at HEASARC. Next, we applied a time filter based on the observed count rate in the 12-15 keV band, and ignored events from periods with high rates in that band. Also events from high-rate modules are ignored. Finally, to search for the pulsed emission from \psr we used only events with energies in the 2.5--10 keV band, the optimum band.
Sometimes the default screening criteria resulted in very small exposure times in which case we used the unfiltered level 1 event files, and followed the same screening steps as outlined before.

\subsection{\nustar\ Observations}

During the 2020-outburst \nustar\ observed \psr\ on 2020 August 5 (ObsID: 80602315002), 2020 August 20 (ObsID: 80602315004), 2020 September 17 (ObsID: 80602315006), and 2020 October 09 (ObsID: 80602315008). We also used 
\nustar\ data of \psr from a $\sim 95$ ks observation taken on 2017 September 17 (ObsID: 40301004002) with the source in quiescent state.

The \nustar\ data were reduced using the standard data reduction process with \nustar\,DAS v1.9.2 and CALDB 20191219. 
We extracted on-source spectra from circular regions of 1$\arcmin$-radius centered at the source position in the 3--70 keV energy band from the \nustar\ observations.  For the spectral fitting, we grouped source spectra using the grppha tool of HEASoft, such that each spectral bin would have a minimum of 25 counts.

\subsection{\swift\ Monitoring}

In this work we also analysed data from the XRT-telescope \citep[][]{Burrows2005} aboard the {\it Neil Gehrels} \swift\ observatory \citep[][]{Gehrels2004},
 operating in Windowed-Timing (WT) mode, providing a time resolution of 1.7675 ms. The \swift\ monitoring campaign of \psr\ started on July 25, 2011 and ended on November 19, 2018. Data from the early phase of the monitoring program up to and including September 4, 2016 has already been used and analysed in \citet{Kuiper2018}. In order to characterize the spin-behavior of \psr\ during its quiescent state before the 2020-outburst we analysed the monitoring data collected between September 17, 2016 and November 19, 2018. More detailed information is given in Sect.
 \ref{sec:timingmodels}.

\section{The August 2020 Outburst} \label{sec:outburst}

On August 1, 2020, the \swift\ telescope reported a short-duration ($\sim 0.1$ s) flare in the BAT detector from \psr \citep{Krimm2020}.  We started ToO observations on August 2 with \nicer\ , obtaining densely sampled observations of the source until November 22, 2020. From our AO2 GO monitoring observations, that started on March 26, 2020, we saw that during the observation performed on June 26, 2020 (MJD 59026) the pulsed count rate in the 2.5-10 keV band had already increased by a factor of $\sim 4$ with respect to the (pre-outburst) quiescent level \citep{Kuiper2020}. The July 25, 2020 (MJD 59055) observation, performed about a week before the magnetar-like burst, showed an even larger, 10 times higher,  2.5-10 keV pulsed count rate than the pre-outburst value. Apparently, the source had already gone into outburst somewhere between May 27, 2020 (MJD 58996; the last AO2 GO observation of the quiescent period) and June 26, 2020.

\begin{figure} 
\vspace{-3.8truecm}
\centering
\includegraphics[width=120mm]{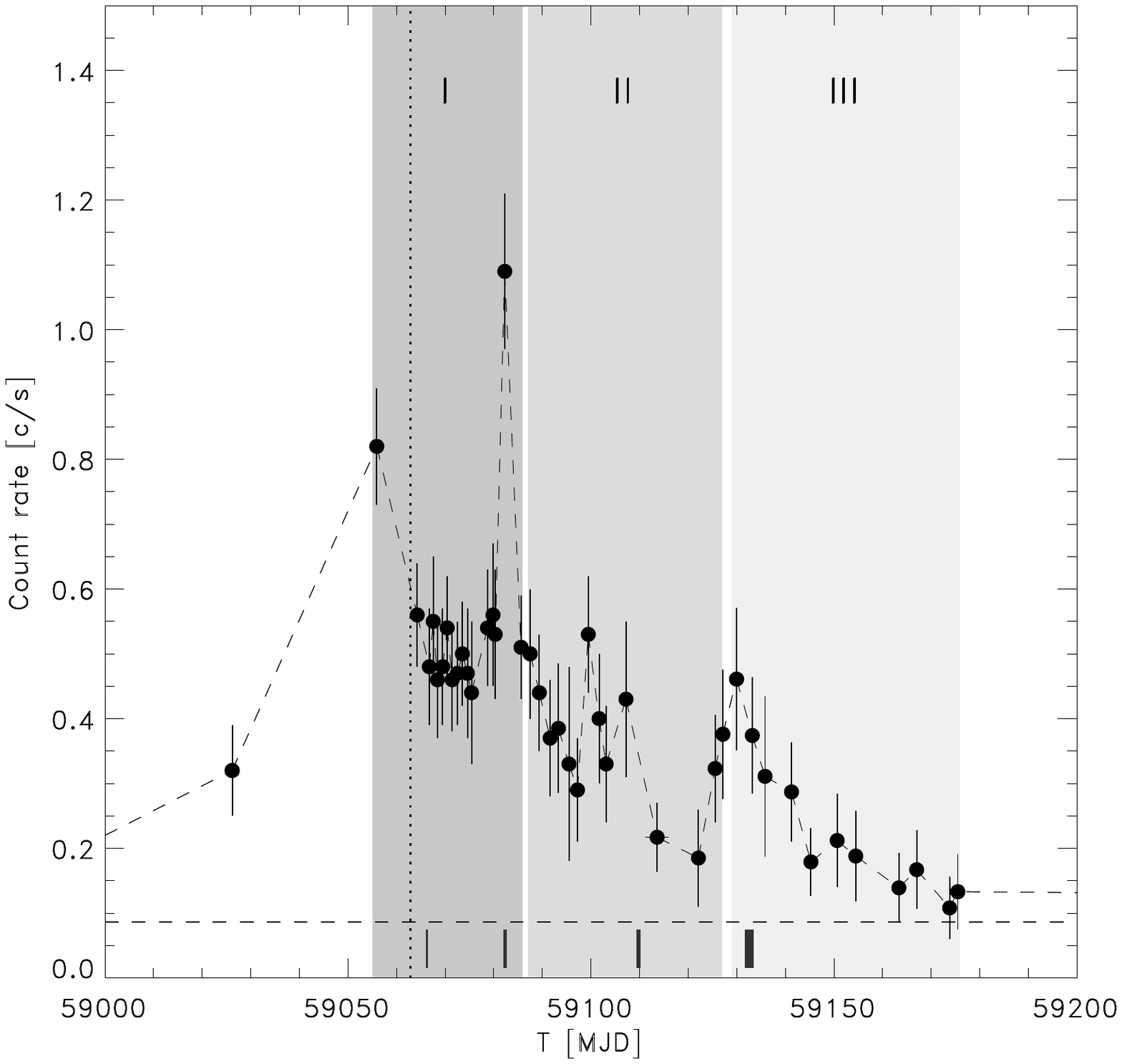}
\vspace{-3.8truecm}
\caption{Pulsed count rate ($2.5–10$ keV) of \psr from \nicer\ data taken between June 26, 2020 and November 22, 
2020. The vertical dotted line indicates the date of the magnetar - like $\sim 0.1$s burst detected by \swift. The horizontal long-dashed line is the quiescent count rate level. The grey areas - indicated by the numbers I, II and III - covering the 2020-outburst era specify the three data periods for which we have phase-coherent timing solutions (see Table \ref{eph_table} entries 5--7 ). The four black bars at the bottom of the plot refer to the four \nustar\ observations performed during the full outburst period.}
\label{fig:countrate}
\end{figure}

Figure \ref{fig:countrate} shows the 2.5 - 10 keV pulsed count rate starting from our AO2 observation on May 27, 2020 through the outburst until the count rate reached near quiescent level on November 22, 2020.  The \swift-BAT burst announced on August 1, 2020 \citep[][]{Krimm2020} is marked by the vertical dotted line, which already occurred after the first recorded rate peaked on July 25, 2020.  During the ToO observations, we recorded at least three more count rate peaks with the highest level reached on August 21, 2020.  The very dense time sampling reveals that the outburst decay was very non-steady with possibly many unresolved variations in flux.  

\cite{Blumer2021} presented radio observations with the Green Bank Telescope (GBT) and X-ray observations with Chandra of \psr and its pulsar wind nebula (PWN) following the 2020 outburst.  Their GBT observations took place August 5 - 6, 2020, only a few days after the \swift\ announcement but actually about ten days after the initial outburst peak flux.  They reported no detectable pulsed radio emission and place an upper limit of 55 mJy.  They also presented results of Chandra ACIS-S observations taken on September 12, 2020, more than one month after the outburst.  Fitting the total (=pulsed plus unpulsed/DC) X-ray spectrum with a power law plus blackbody, they reported a softer power law index of $\Gamma = 1.7 \pm 0.3$ compared to the quiescent value of $\Gamma =1.2 \pm 0.1$ and the appearance of a thermal component with temperature $kT = 0.7 \pm 0.1$ keV.  Their measured 0.5 - 10 keV flux was 2.4 times the quiescent flux level, consistent with our \nicer\ count rate on that date (see Fig. \ref{fig:countrate}).  They detected no significant increase in X-ray flux or spectral index of the PWN.  \citet{Livingstone2011} also found no change in X-ray flux or spectrum of the PWN following the 2006 outburst.
\begin{table*}[t]
\caption{Phase-coherent ephemerides for \psra\ as derived from \nicer-XTI and \swift-XRT (monitoring) data covering the time period
MJD 57871-59532 (April 28, 2017 -- November 14, 2021).}
\label{eph_table}
\begin{center}
\begin{tabular}{lccclllccc}
\hline
Entry$^{\ast}$ &  Start &  End  &   t$_0$, Epoch   & \multicolumn{1}{c}{$\nu$}   & \multicolumn{1}{c}{$\dot\nu$}      & \multicolumn{1}{c}{$\ddot\nu$}       & \multicolumn{1}{c}{$\dddot\nu$}           & RMS$^\dagger$ & $\Phi_{0}^\ddagger$  \\
 \#   &  [MJD] & [MJD] &     [MJD,TDB]    & \multicolumn{1}{c}{[Hz]}    & \multicolumn{1}{c}{[$10^{-11}$ Hz/s]}  & \multicolumn{1}{c}{[$10^{-21}$ Hz/s$^2$]} & \multicolumn{1}{c}{[$10^{-24}$ Hz/s$^3$]} &      &                \\
\hline
\multicolumn{10}{l}{{\dotfill} Pre-Outburst 2020 {\dotfill}}\\
1           & 57871 & 58061 & 57956.0     & 3.044\,250\,906(4)   & -6.636\,32(7)             &   4.63(79)        & 0.0       & 0.044    & 0.4167\\
2           & 58167 & 58422 & 58402.0     & 3.041\,697\,558(20)  & -6.619\,07(50)            &   3.67(54)      & 0.0       & 0.025    & 0.4258\\
3           & 58261 & 58671 & 58312.0     & 3.042\,212\,358(3)   & -6.621\,40(4)             &   2.93(3)       & 0.0       & 0.029    & 0.1074\\
4           & 58698 & 58996 & 58893.0     & 3.038\,892\,369(2)   & -6.604\,60(3)             &   5.72(9)       & 0.0       & 0.042    & 0.9539\\
\multicolumn{10}{l}{{\dotfill} Outburst 2020 {\dotfill}}\\
\vspace{-0.25cm}\\
5           & 59055 & 59086 & 59064.0     & 3.037\,919\,719(23)  & -6.725\,6(25)             &  -2659(45)      & 0.0       & 0.038    & 0.8216\\
6           & 59087 & 59129 & 59099.0     & 3.037\,707\,087(50)  & -7.018\,0(52)             &   1776(98)      &-1.37(14)  & 0.026    & 0.1456\\
7           & 59129 & 59175 & 59145.0     & 3.037\,432\,343(15)  & -6.849\,8(12)             &    298(22)      & 0.0       & 0.038    & 0.4163\\
\multicolumn{10}{l}{{\dotfill} Post-Outburst 2020 {\dotfill}}\\
\vspace{-0.25cm}\\
8           & 59281 & 59418 & 59318.0     & 3.036\,433\,103(8)   & -6.709\,6(2)              &   5.6(1.0)      & 0.0       & 0.022    & 0.0535\\
9           & 59434 & 59532 & 59523.0     & 3.035\,250\,253(16)  & -6.667\,2(4)              &   0.0           & 0.0       & 0.018    & 0.7372\\
\hline
\multicolumn{10}{l}{$^{\ast}$ Entries 1-2 are solely \swift-XRT ToA based covering April 28 -- November 4, 2017 (entry-1) and February 18 -- November 19, 2018 (entry-2),}\\ 
\multicolumn{10}{l}{entry 3 is based on a combination of \swift-XRT and \nicer\ ToA's starting at the first \nicer\ observation performed on}\\ 
\multicolumn{10}{l}{May 23, 2018 (MJD 58261) and entries 4--9 are solely \nicer\ TOA based. }\\ 
\multicolumn{10}{l}{$^{\ast}$ Solar System planetary ephemeris DE200 has been used in the barycentering process.}\\
\multicolumn{10}{l}{$^{\dagger}$ The RMS is the root-mean-square of the deviation from the best fit in phase units.}\\
\multicolumn{10}{l}{$^{\ddagger}$ $\Phi_{0}$ is the phase offset to apply in the phase calculation to obtain consistent alignment \citep[see e.g. Eq. 1 in][]{Kuiper2018}. }\\
\end{tabular}
\end{center}
\end{table*}

\subsection{Timing analysis}  \label{sec:timing}
In this work, constituting an extension of the timing analyses presented in earlier work by \citet{Kuiper2009} and \citet{Kuiper2018}, we performed timing analyses of the pre-outburst, outburst and post-outburst episodes of the 2020-outburst using data from \nicer- and (for a part of the pre-outburst period, 2017-2018) \swift-monitoring observations to make comparisons between quiescent- and outburst states possible. \swift\ and \nicer\ monitoring overlapped between May--November 2018 enabling cross-checking and verification of the timing characteristics.

These analyses yielded phase-coherent timing models (ephemerides) in terms of three, sometimes four, timing parameters specifying the frequency $\nu$, frequency derivative $\dot\nu$, frequency second $\ddot\nu$ and third order $\dddot\nu$ derivatives
evaluated at a certain epoch t$_0$. These models accurately describe the rotation behavior of the pulsar as a function of time, and so tracking the spin-evolution of the source. 

Pulse-phase folding timing data from any instrument using these ephemerides, adopting different energy selections, yielded energy-resolved pulse-phase distributions, often called pulse-profiles.
From these the pulse shape as a function of energy can be determined for any emission state of the pulsar.

The models also enable us to separate at any epoch the pulsed emission component from the total emission which has contributions from the surrounding PWN and SNR, and the pulsar itself either pulsed or DC. So, we can study the pulsed flux (or count rate) as a function of time, probing the outburst. 

\begin{figure}[t] 
\vspace{-2.5truecm}
\includegraphics[width=92mm]{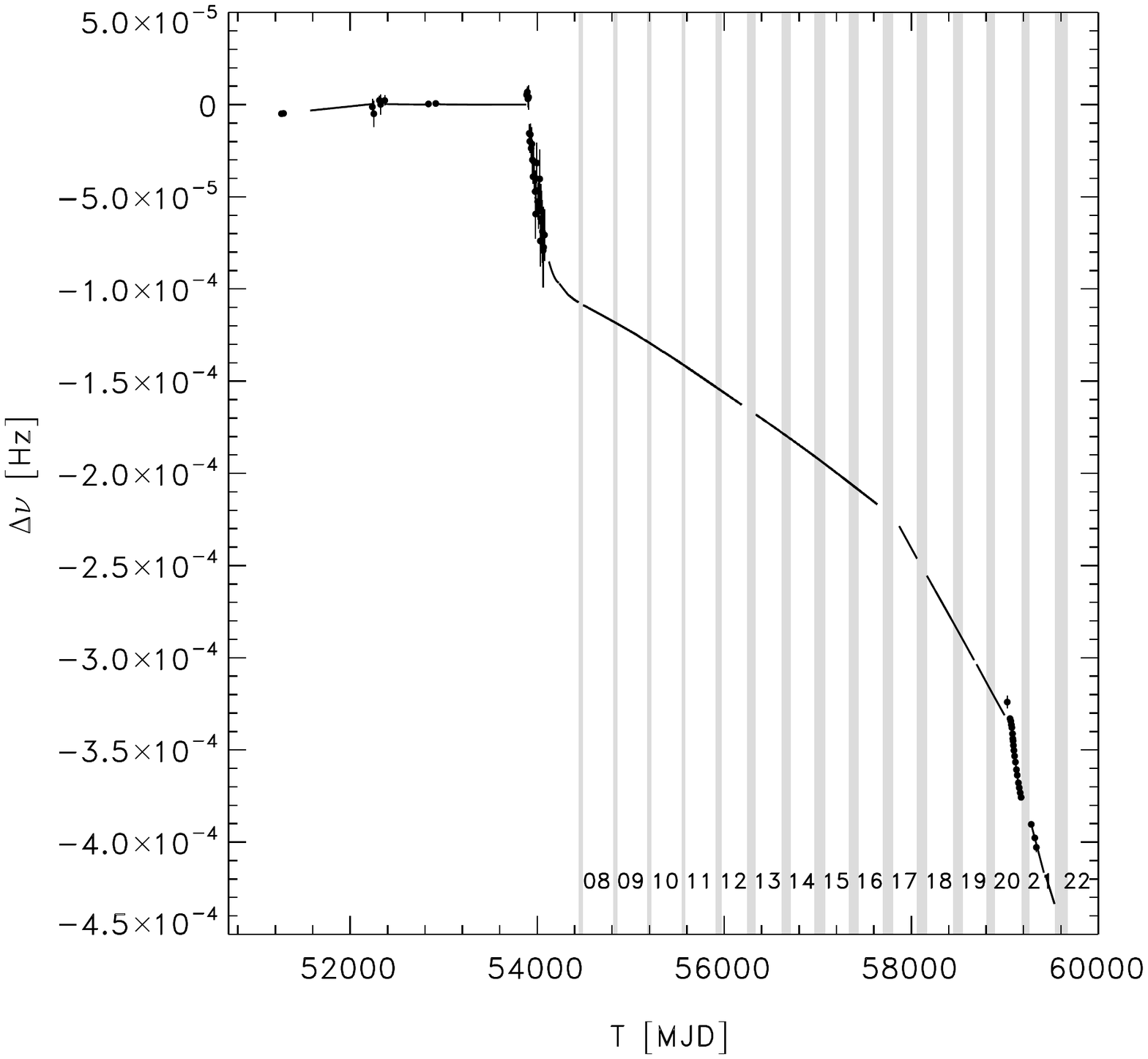}\includegraphics[width=92mm]{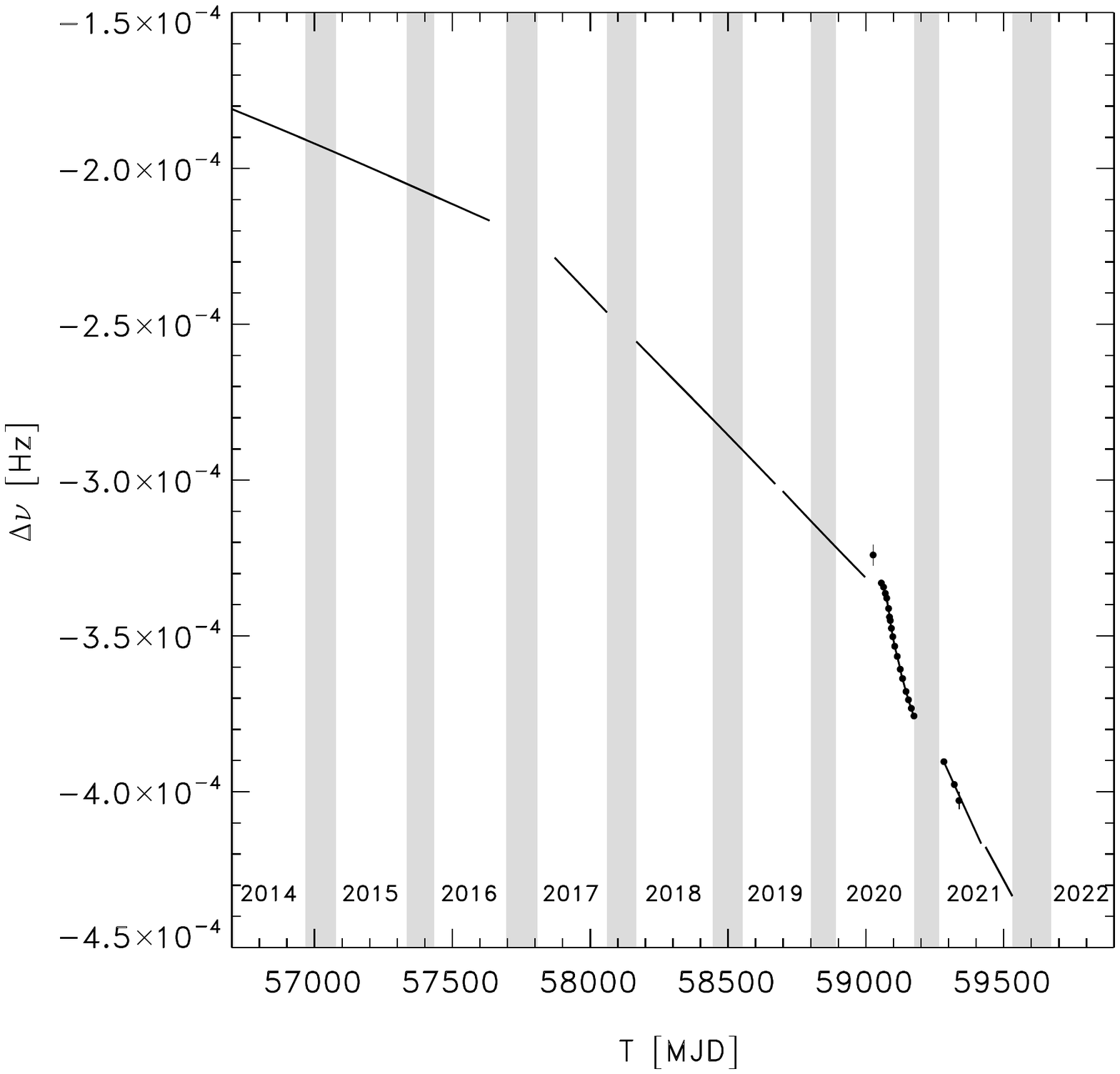}
\vspace{-2.7truecm}
\caption{Left: Spin frequency evolution $\Delta\nu(t)$ of \psr \citep[relative to the pre-2006 outburst ephemeris, see e.g.][]{Kuiper2015,Gavriil2008,Livingstone2010,Livingstone2011} since its discovery in \rxte-PCA data from April 1999 up to
November 2021. The 2006 and 2020 glitches are clearly visible near MJD 54000 and MJD 59000, as well as an (accelerated) spin-down rate feature/glitch that pops up around MJD 57700 near/during the \swift\ 2016--2017 observation gap.  Right: Zoom in of the rotation frequency evolution during the period of \swift-XRT and \nicer\ timing from early 2014 until November 2021. \swift\ and \nicer\ data gaps are shown as grey hatched vertical areas.}
\label{fig:deltanu}
\end{figure}

\subsubsection{Timing models (ephemerides)}  \label{sec:timingmodels}
The first step in the timing analysis is the conversion of spacecraft arrival times (usually evaluated in TT or TDT time system) to solar-system barycentered times (TDB time frame) using the instantaneous spacecraft ephemeris (specifying the position and velocity of the spacecraft), solar-system ephemeris (in this work JPL-DE200), a \cxo\ X-ray based accurate location of the pulsar \citep{Helfand2003} and for \swift\ up-to-date clock-correction information.
We performed this barycentering process for \swift\ - XRT (WT-mode) observations 00032031149 -- 00032031216 executed between September 15, 2016 and November 19, 2018 (MJD 57646 -- 58441). An equivalent (IDL-based) barycentering procedure was used for the \nicer\ observations.

From the barycentered \swift\ and \nicer\ event lists of each observation (or combination of observations if the exposure of a single observation was too low) we derived so-called pulse arrival times (ToA's). For \swift-XRT (2.5--10 keV) on average we typically needed 15-20 ks net exposure time to detect the timing signal of \psr\ above $3\sigma$ in a $25''$ extraction area centered on the source \citep[see e.g. Sect. 3.1.2 of][]{Kuiper2018}. For \nicer\ (2.5--10 keV) the required net exposure is typically 5--8 ks to reach a significant pulsed signal detection depending on the instantaneous background rate.

From these ToA's we constructed phase-coherent timing models covering a $\sim 4.5$ year period from April 28, 2017 till November 14, 2021 according to the method outlined in Sect. 4 of \citet{Kuiper2009}. The results are listed in Table \ref{eph_table} and consist of eight separate models: four covering the pre-outburst episode (entries 1--4 of which 1--2 are solely based on \swift\ monitoring), three covering the (densely sampled) outburst period (entries 5--7), and finally two covering the post-outburst era (entries 8--9).
The spin-evolution of \psr\ since its discovery in April 1999 with respect to the pre-2006 outburst ephemeris is shown in the left panel of Fig. \ref{fig:deltanu}, while the right panel zooms in on the time period analysed in this work. In the latter period from early September 2016 till mid November 2021 we discovered several timing discontinuities in both $\nu$ and $\dot\nu$ which are summarized below:

\newcounter{sumlist}
\begin{list}%
{(\roman{sumlist})}{\usecounter{sumlist}\setlength{\rightmargin}{\leftmargin}\itemsep=0pt}
\item MJD 57635--57871 (September 4, 2016 - April 28, 2017): A clear glitch in $\dot\nu$ showing an acceleration in the spin-down rate occuring just before or in the 2016--2017 \swift\ monitoring data gap.
\item MJD 58061--58167 (November 4, 2017 - February 18, 2018): Phase-coherence loss during the 2017--2018 \swift\ monitoring data gap.
\item MJD 58671--58698 (July 7, 2019 - August 3, 2019):  A small glitch of fractional size $\Delta\nu_g/\nu_g = (7.3\pm 0.4)\times 10^{-8}$ occurred during the \nicer\ monitoring. In spite of its small amplitude, the glitch could be well characterized because of the densely sampled \nicer\ ToO-observations in the post-glitch period. 
Changes in both $\dot\nu$ and $\ddot\nu$ were also detected, with values $\Delta \dot\nu{_g}/ \dot\nu{_g} = (4.5\pm 0.3)\times 10^{-4}$ and $\Delta \ddot\nu_g= (2.66\pm 0.13)\times 10^{-21}$ Hz/s$^2$, all evaluated at an assumed glitch epoch $t_g$ of MJD 58685. This timing glitch could easily be distinguished from red noise. 
\item MJD 58996--59055 (May 27, 2020 - July 25, 2020): A strong glitch of fractional size $\Delta\nu_g/\nu_g \sim 3\times 10^{-6}$, comparable in size to the 2006 outburst \citep[see Sect.4.2 of][for more details]{Kuiper2009}, heralded very likely the start of the 2020 outburst of \psr.
\item MJD 59175--59281 (November 11, 2020 - March 8, 2021): Phase-coherence loss during the \nicer\ 2020--2021 data gap of $\sim 3.4$ month duration due to one or more glitches.
\item MJD 59418--59434 (July 23, 2021 - August 8, 2021): A small glitch of fractional size $\Delta\nu_g/\nu_g \sim (0.5-0.66)\times 10^{-6}$ occured during regular monthly \nicer\ monitoring.
\end{list}

A linear fit to the spin-down rates in Table \ref{eph_table} gives an indication of the effective evolution of this quantity on time, with $\ddot\nu=3.9(3)\times 10^{-21}$  Hz/s$^2$. 
This indicates a braking index before the outburst (\swift\ + \nicer\ data) of $n=2.7\pm 0.2$.
Interestingly, the braking index after the 2006 outburst was seen to decrease from $2.65\pm 0.01$ to $2.16 \pm 0.13$ \citep{Livingstone2011}.
Our preliminary measurement indicates that between 2006 and 2020 the rotation somehow went back to its original pre-2006-outburst evolution. 
There is still not enough data to measure the braking index after the 2020 outburst.

\begin{figure}[t]
\vspace{-2.25truecm}
\includegraphics[angle=-180,width=\textwidth]{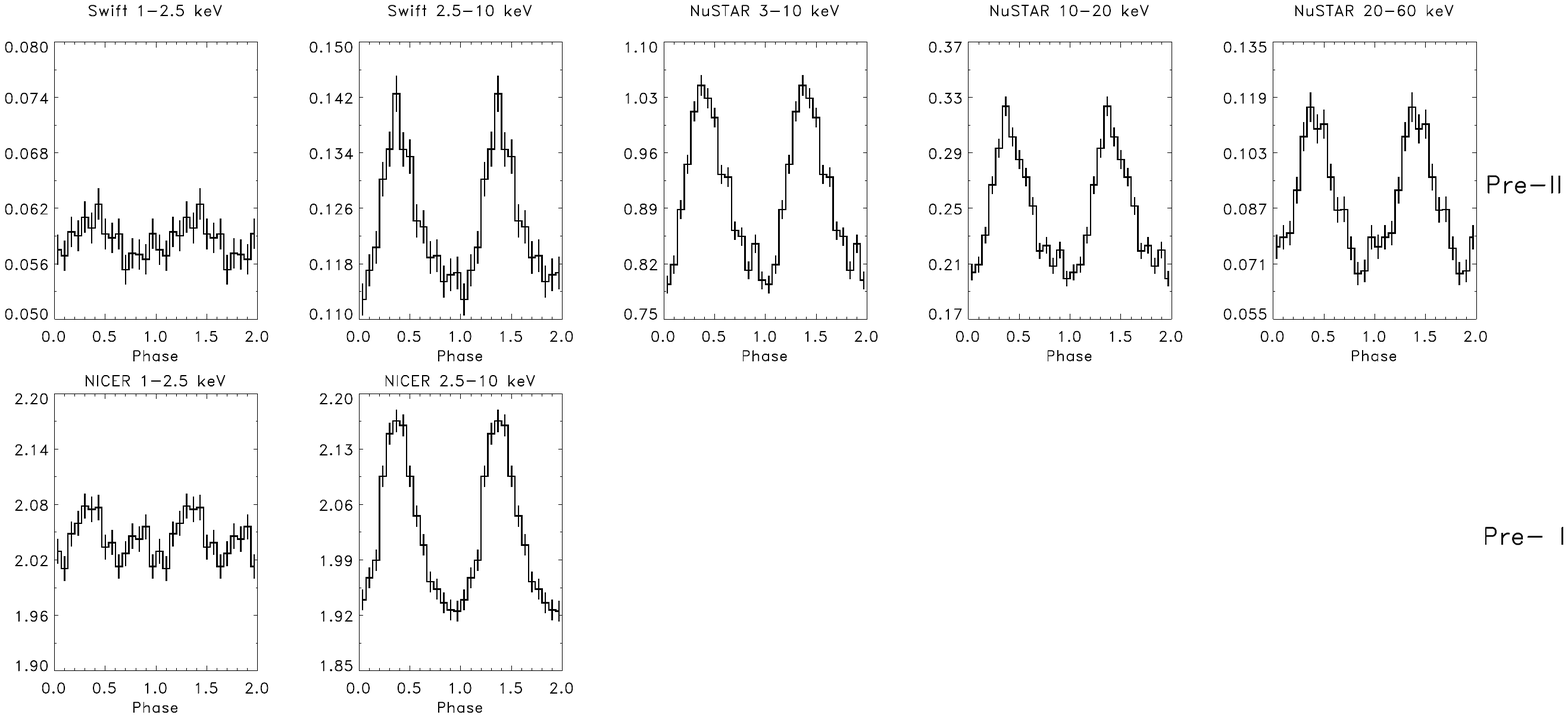}
\vspace{-2.5truecm}
\caption{Pulse profiles of \psr for different stages prior to the 2020-outburst, indicated by Pre-II (\swift\ 2017--2018; MJD 57871--58442; April 28, 2017 -- November 20, 2018 totaling an exposure time of 319.4 ks, \nustar\ September 17--20, 2017; exposure time $\sim 95$ ks) and Pre-I (\nicer\ 2018 -- 2020 pre-outburst; MJD 58261--58996; May 23, 2018 -- May 27, 2020; exposure time $\sim 166.7$ ks), for different energy bands as observed by \swift, \nicer\ and \nustar. Along the y-axis the count rate (in cts/s) is shown. Error bars are at $1\sigma$ confidence.}
\label{fig:pre_outburst}
\end{figure}

\subsubsection{Pulse profile evolution}  \label{sec:profiles}
\begin{figure}[t] 
\includegraphics[angle=-180,width=\textwidth]{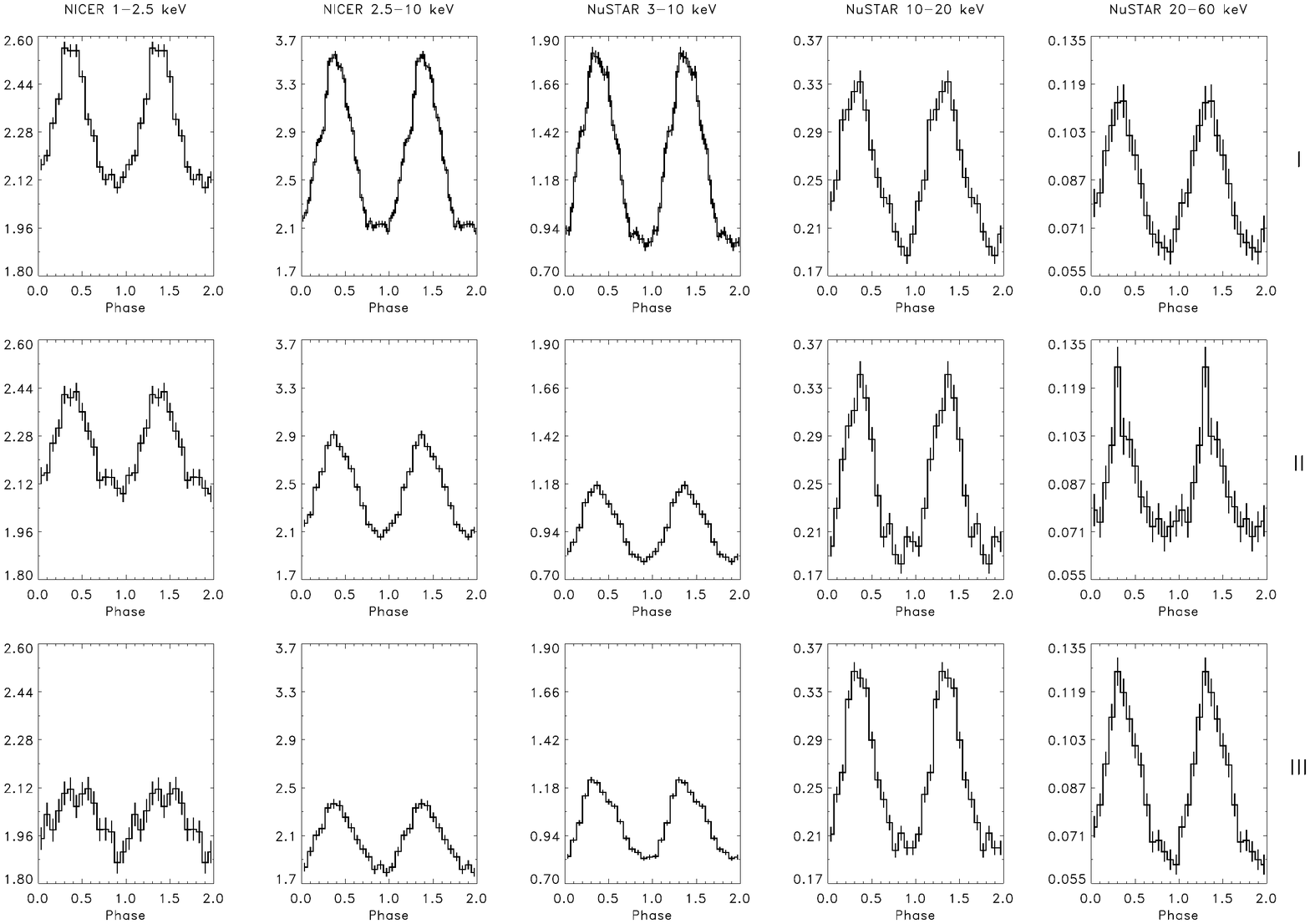}
\vspace{-1truecm}
\caption{Pulse profiles of \psr during different phases of the 2020-outburst, indicated by I (MJD 59055--59086; July 25 -- August 24, 2020), II (MJD 59087--59127; August 26 -- October 5, 2020) and III (MJD 59129--59175; October 7 -- November 22, 2020, for different energy bands as observed by \nicer\ and \nustar.
\nustar\ observed \psr on August 5 and August 20, 2020, during segment-I, on September 17, 2020 during II, and October 9, 2020 during III. Along the y-axis the count rate (in cts/s) is shown. Error bars are at $1\sigma$ confidence. The columns show the pulse profiles for a certain energy band for either \nicer\ or \nustar\ with a fixed range in ordinate making comparisons between profiles from different outburst phases more comfortable. Significant changes in shape of the pulse phase distributions and intensity of the pulsed emission are only detectable below $\la 10$ keV.}
\label{fig:outburst}
\end{figure}

\begin{figure}[t] 
\vspace{-2.75truecm}
\centering
\includegraphics[trim= 144 144 144 144,width=0.4\textwidth]{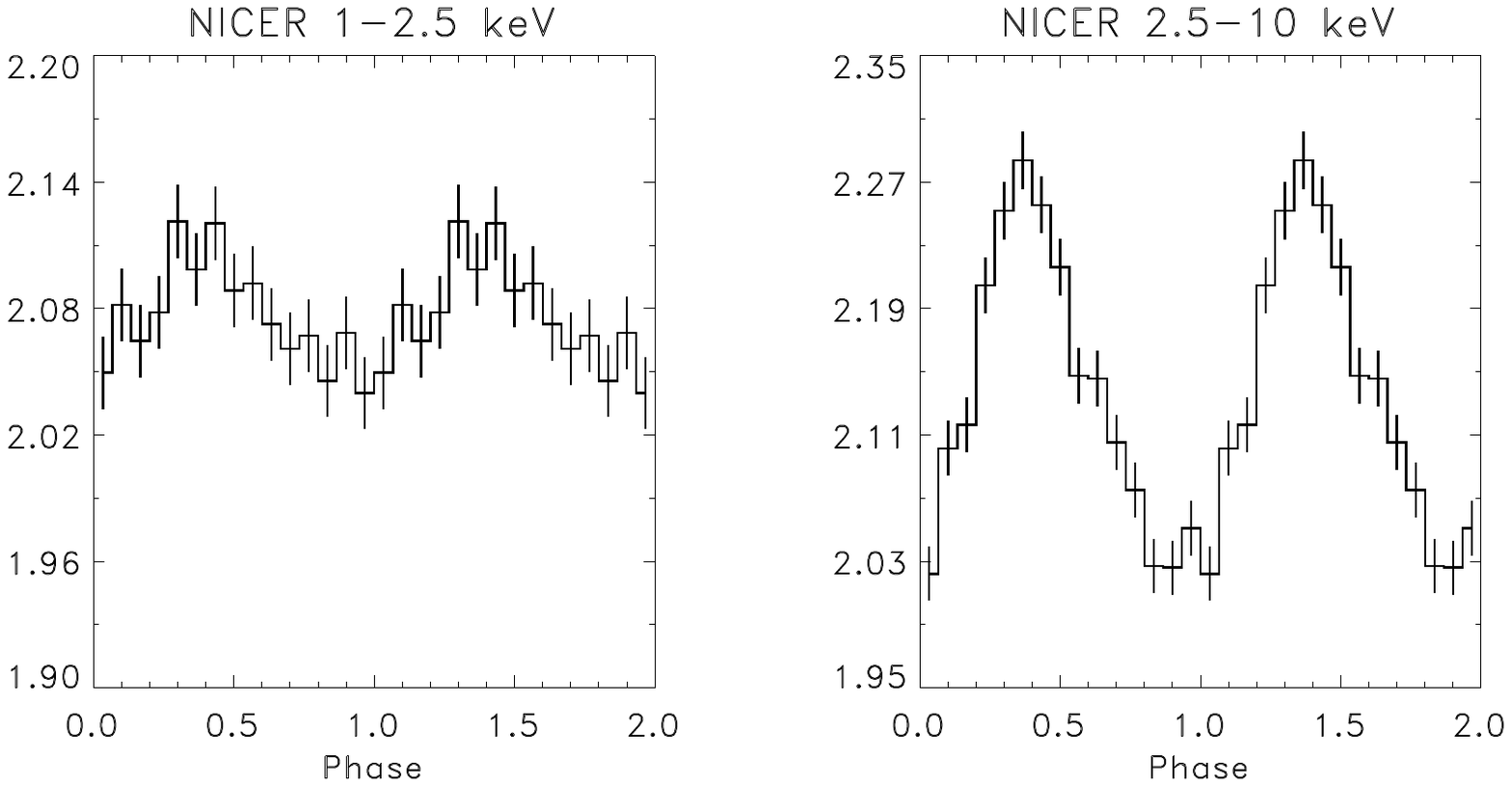}
\vspace{-3.5truecm}
\caption{Pulse profiles of \psr for the Post-outburst 2020 period (MJD 59281--59531; March 8 -- November 13, 2021) as observed by \nicer\ in two different energy bands. Along the y-axis the count rate (in cts/s) is shown. Error bars are at $1\sigma$.}
\label{fig:post_outburst}
\end{figure}

We applied the timing models listed in Table \ref{eph_table} in a pulse-phase folding procedure to obtain pulse-phase distributions for several different energy bands for the pre-outburst period (split into two parts related to \swift\ and \nicer\ monitoring), for the outburst episode subdivided into three parts driven by ephemeris validity constraints, and finally for the post-outburst era. The resulting phase-distributions are shown in Figs. \ref{fig:pre_outburst}, \ref{fig:outburst} and \ref{fig:post_outburst} for the pre-outburst, outburst and post-outburst periods, respectively. 

The pre-outburst period running from April 28, 2017 to May 27, 2020 was subdivided into two parts with \swift\ (Pre-II; MJD 57871--58442) and \nicer\ (Pre-I; MJD 58261--58996) monitoring.
Pulse phase folding 51 \swift\ observations, totaling 319.402 ks exposure time, resulted in the phase distributions shown in the upper-left two frames of Fig. \ref{fig:pre_outburst} with $Z_2^2$ significances of $2.8\sigma$ and $13.0\sigma$ for the 1--2.5 and 2.5--10 keV bands, respectively. During the \swift\ monitor period also \nustar\ observed \psr during September 17--20, 2017 for $\sim 95$ ks while the source was in quiescent state. The well-known pulse-shape \citep[see e.g. Fig. 3 of][]{Kuiper2009} is clearly visible up to $\sim 60$ keV in these \nustar\ data (see upper right three frames of Fig. \ref{fig:pre_outburst}). The second part of the pre-outburst period (Pre-I) referred to solely \nicer\ monitoring observations comprising 166.67 ks of (screened) exposure time. The corresponding pulse-phase distributions are shown in the lower-left frames of Fig. \ref{fig:pre_outburst} for the 1--2.5 ($Z_2^2$ significance of $4.8\sigma$) and 2.5--10 keV bands.

The post-outburst period from March 8 -- November 13, 2021 (MJD 59281--59531) using merely \nicer\ monitoring observations, totalling 103.55 ks of screened exposure time, shows pulse-phase distributions for the 1--2.5 ($Z_2^2$ significance of $4.1\sigma$) and 2.5--10 keV bands (see Fig. \ref{fig:post_outburst}) equivalent in shape and strength to the corresponding \nicer\ pre-outburst distributions. 

It is clear from the pre- and post-outburst observations, representing the quiescent state, that the pulsed emission in the 1--2.5 keV band is weak, mainly coming from events with measured energies in the range 2--2.5 keV.

The situation changed dramatically during the outburst episode (July 25 -- November 22, 2020; MJD 59055-59175) that was split into three smaller time segments, indicated by I, II and III (\nicer\ screened exposures 86.014, 40.694 and 20.064 ks, respectively), driven by phase-coherency\footnote{During a regular \nicer\ monitoring observation on June 26, 2020 (MJD 59026) the source was already in outburst, however, we could not phase-connect its timing data} (see Table \ref{eph_table}). 

Figure \ref{fig:outburst} shows the evolution of the pulse morphology and strength for different energy bands during different stages of the outburst as observed by \nicer\ and \nustar.
The most left column showing the \nicer\ 1--2.5 keV lightcurves demonstrates the emergence of a very strong (soft) pulsed emission component which gradually fades away towards its quiescent level. The pulsed significance $Z_2^2$ in the 1--2.5 keV band evolves from $31.5\sigma$ to $15.2\sigma$ to $6.0\sigma$ for the three segments I, II and III, respectively.

Both the \nicer\ 2.5--10 keV and \nustar\ 3--10 keV bands show a similar fading evolution, however, contrary to the quiescent pulse morphology a prominent `shoulder' or bump appears during segment I observations near phase $\sim 0.2$ before primary maximum at phase 0.37. This feature also fades away in course of the outburst. It is also clear from the 1--2.5, 2.5--10 and 3--10 keV pulse-phase distributions that an underlying enhanced DC-component appeared that fades away across the outburst.

On the contrary, the pulsed emission above 10 keV as shown by \nustar\ in the two right most columns of Fig. \ref{fig:outburst} seems to not vary dramatically and is comparable to its quiescent state properties (c.f. Fig.\ref{fig:pre_outburst}; Pre-II - right two panels).

\subsubsection{Pulsed flux evolution during the 2020-outburst}  \label{sec:fluxoutburst}
From the pulse-phase distributions per observation (or combination of observations)
pulsed excess counts can be determined for any energy band by estimating the excess counts above the DC or unpulsed level evaluated from the 0.8--1.05 phase window \citep[off-pulse phase; see e.g. Fig. 3 of][]{Kuiper2009}.
These pulsed excess counts are subsequently converted to rates using the (screened) exposure times of the involved observation(s). The 2.5--10 keV pulsed count-rate across the 2020-outburst episode as observed by \nicer\ is shown in Fig. \ref{fig:countrate}.
The different segments - I, II and III - are hatched while solid vertical bars at the bottom of the plot indicate the performed \nustar\ observations. The horizontal dashed line represents the quiescent 2.5--10 keV level and the vertical dotted line indicates the \swift\ detection of the magnetar-like $\sim 0.1$s burst \citep{Krimm2020}, triggering our \nicer\ ToO program. It is clear from this plot that \psr was already in outburst before the occurrence of the magnetar-like burst. Several re-flarings are also visible in Fig. \ref{fig:countrate} in particular the one on August 21, 2020 (MJD 59082). Initially, we suspected that this was a statistical fluke, however, a deeper study reveals that this enhancement of about 1 hour duration was also clearly detectable in a contemporaneous \nustar\ observation (ObsID: 80602315004), and so demonstrates the validity of the event. The feature mimics a mini-outburst, for the first time introduced by \citet{Kuiper2012} (see their Sect. 3.1.1) in the study of the high-energy characteristics of the 2008 and 2009-outbursts of magnetar \eins.

The quiescent state level is reached near MJD 59175 (November 22, 2020), the last \nicer\ observation before the 2020--2021 data gap.

\subsection{Spectral Evolution}  \label{sec:spectra}
We carried out spectral fits on the pulsed component only, meaning that the photons collected during off-pulse phases were treated as background and subtracted from the on-pulse spectra. To examine the broadband spectral behavior of the pulsed emission, we first jointly fit \nicer\ and \nustar\ spectra at four different epochs. We used the tbabs model with corresponding interstellar medium abundance as described in \citet{WilmsAM2000} to determine the hydrogen column density ($N_{\scriptsize{\textrm{H}}}$) and linked this parameter across all data sets. The spectra were fit with a two-component model consisting of a hard power law and a soft blackbody. The parameters of these two components are linked between data sets obtained at the same epoch. An additional constant component was introduced to account for the cross-calibration between instruments and the flux difference between data sets. 

\begin{figure} 
\centering
\includegraphics[width=150mm]{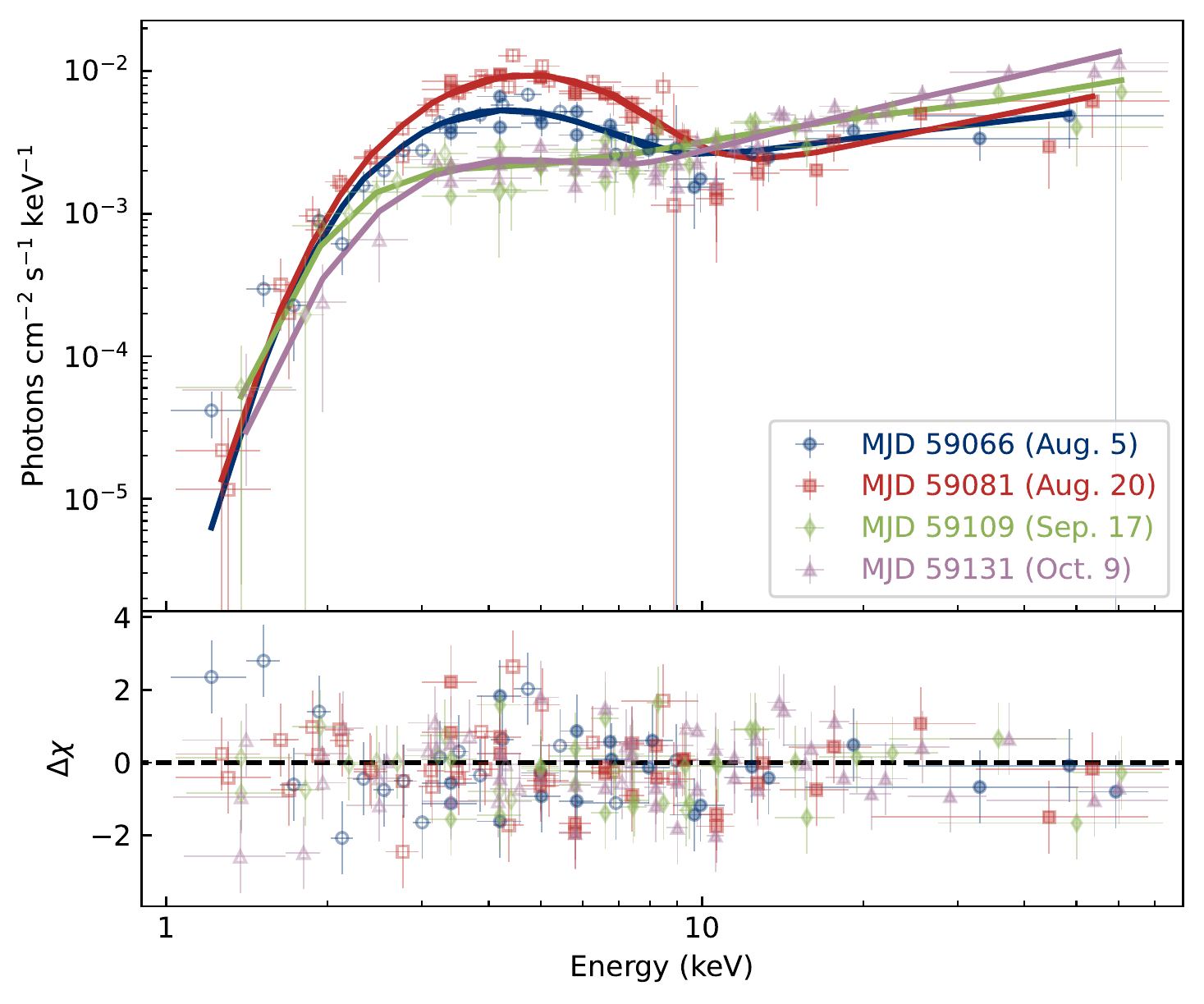}
\caption{Spectra extracted from \nicer\ data (open points) and \nustar\ data (closed points). The solid lines indicate the best-fitting model to data with the corresponding color. The bottom panel shows the corresponding residuals in terms of $\sigma$.}
\label{fig:specevolution_nustar}
\end{figure}

The first \nustar\ observation (ObsID 80602315002) was made together with \nicer\ ObsID 3033290103 on MJD 59066, and we fit these two data sets by linking all the spectral parameters. The other three \nustar\ observations did not have corresponding \nicer\ observations on the same dates, so we performed simultaneous fits with nearby \nicer\ observations that bracketed the \nustar\ ones. For example, we jointly fit the \nustar\ observation on MJD 59081 (ObsID 80602315004) with \nicer\ observations on MJD 59080 and 59082. Similar work was done for observations on MJD 59109 (ObsID 80602315006) and MJD 59131 (ObsID 80602315008).  The results are shown in Table \ref{tab:spec_joint_nustar_nicer} and the unfolded spectra of these four data sets are shown in Figure \ref{fig:specevolution_nustar}.  The $N_{\textrm{H}}$ was constrained to be $(6.1\pm0.8)\times10^{22}$ cm$^{-2}$. The photon indices of these four data sets varied slightly but remained consistent at around $\Gamma\approx1.2$. On the other hand, the blackbody component varied significantly, indicating that the flux variability during the outburst is dominated by changes in the blackbody component. This can also be seen from the pulse profiles as a function of energy.

\begin{figure} 
\centering
\includegraphics[width=150mm]{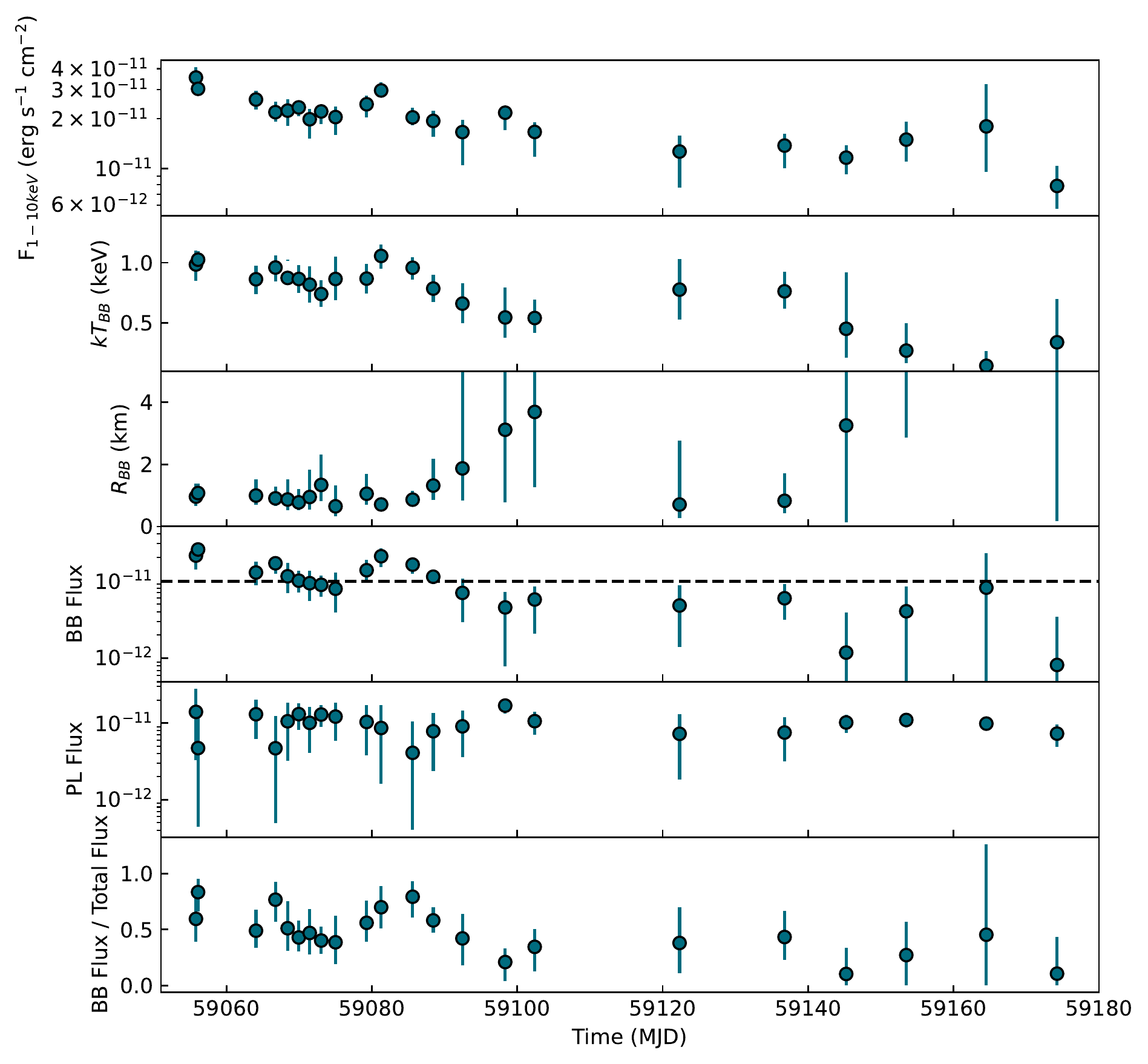}
\caption{Spectral evolution of the pulsed emission of \psr observed with \nicer\ following the outburst on 2020 August 01. From upper to lower, panels show the evolution of the unabsorbed flux in the 1--10 keV band (in units of erg~cm$^{-2}$~s$^{-1}$), the blackbody temperature (in units of keV), the emitting radius (assuming a distance of 6.5 kpc), the BB flux, PL flux, and the ratio of BB flux to the total flux. The horizontal dashed line denotes the mean value of the PL flux across all data sets. \label{fig:specevolution}}
\end{figure}

We then traced the spectral evolution in the soft X-ray band using all \nicer\ data sets. We tried to fit individual data sets by freezing the $N_{\textrm{H}}$ at $6.1\times10^{22}$ cm$^{-2}$ and allowing all the blackbody and power law parameters to be free. However, the \nicer\ energy range is not sensitive to the hard power law component, and the spectral parameters could not be well constrained. Therefore, we froze the $\Gamma$ at 1.2 and allowed the $kT_{\textrm{BB}}$, $R_{\textrm{BB}}$, and power-law normalization to be free parameters in the following analysis. Additionally, the number of X-ray photons decreased as the flux decreased and the exposure time of each observation was different. Therefore, we jointly fit one to three data sets by linking the aforementioned three spectral parameters. The resulting unabsorbed fluxes in the 1-10 keV band, fluxes of individual components in the same energy band, and three spectral parameters are shown in Figure \ref{fig:specevolution}, and the parameters are shown in Table \ref{tab:spec_nicer}. The overall flux, $kT_{\textrm{BB}}$, and blackbody flux decreased over time. A mini flare on MJD 59082 (August 21) is observed. During the flare, the $kT_{\textrm{BB}}$ and flux increased. On the other hand, the power law flux remained constant throughout the entire observation run, suggesting that the outburst is dominated by the increase of the surface emission. A few days after the flare, the blackbody component could not be well constrained and the power law dominated the spectrum in the soft X-ray band.

\begin{deluxetable*}{lllccc}[t]
\tablecaption{Best-Fit parameters of joint \nustar\ -- \nicer\ observations \label{tab:spec_joint_nustar_nicer}} 
\tablehead{\colhead{Time (MJD)} & \colhead {NuSTAR ObsID} & \colhead {NICER ObsID} & \colhead{$\Gamma$} & \colhead{$kT_{\textrm{BB}}$ (keV)} & \colhead{$R_{\textrm{BB}}$ (km)}}
\startdata
59066 & 80602315002 & 3033290103 & $1.4_{-0.3}^{+0.5}$ & $0.93\pm0.08$ & $0.6\pm0.3$\\
59081 & 80602315004 & 3033290115,116 & $1.3_{-0.5}^{+0.3}$& $1.01_{-0.05}^{+0.06}$ & $0.9_{-0.2}^{+0.3}$\\
59109 & 80602315006 & 3033290126,127 & $1.4\pm0.2$ & $0.5_{-0.1}^{+0.2}$ & $4_{-3}^{+14}$\\
59131 & 80602315008 & 3033290132,135,136 & $1.1\pm0.1$ & $0.8\pm0.1$ & $0.6_{-0.3}^{+1.2}$\\
\enddata
\end{deluxetable*}

\begin{deluxetable*}{llcccc}
\tablecaption{Best-Fit parameters of \nicer\ observations \label{tab:spec_nicer}} 
\tablehead{\colhead {Time} & \colhead {NICER ObsID} & \colhead{$kT_{\textrm{BB}}$ } & \colhead{$R_{\textrm{BB}}$} & \colhead {Flux} & \colhead {BB Flux} \\
\colhead{(MJD)} & & \colhead{(keV)} & \colhead{(km)} & \colhead{($10^{-11}$ erg s$^{-1}$ cm$^{-2}$)} & \colhead{($10^{-11}$ erg s$^{-1}$ cm$^{-2}$)} }
\startdata
59056 & 3598010801 & $1.0\pm0.1$ & $1.0_{-0.3}^{+0.4}$ & $3.5_{-0.4}^{+0.5}$ & $2.1_{-0.7}^{+0.8}$\\
59056 & 3598010802 & $1.03_{-0.09}^{+0.07}$ & $1.1_{-0.2}^{+0.3}$ & $3.0_{-0.3}^{+0.4}$ & $2.5_{-0.5}^{+0.4}$\\
59064 & 3033290101,102 & $0.9\pm0.1$ & $1.0_{-0.3}^{+0.5}$ & $2.6\pm0.3$ & $1.3_{-0.4}^{+0.5}$\\
59067 & 3033290103 & $1.0\pm0.1$ & $0.9_{-0.3}^{+0.4}$ & $2.2\pm0.3$ & $1.7_{-0.4}^{+0.3}$\\
59068 & 3033290105 & $0.9\pm0.1$ & $0.9_{-0.3}^{+0.6}$ & $2.2\pm0.4$ & $1.1\pm0.5$\\
59070 & 3033290106,107 & $0.9\pm0.1$ & $0.8_{-0.3}^{+0.4}$ & $2.3_{-0.3}^{+0.2}$ & $1.0_{-0.3}^{+0.4}$\\
59072 & 3033290108 & $0.8\pm0.2$ & $0.9_{-0.4}^{+0.9}$ & $2.0_{-0.5}^{+0.3}$ & $0.9\pm0.4$\\
59073 & 3033290109,110 & $0.7\pm0.1$ & $1.3_{-0.5}^{+0.9}$ & $2.2_{-0.4}^{+0.2}$ & $0.9\pm0.3$\\
59075 & 3033290111,112 & $0.9\pm0.2$ & $0.6_{-0.3}^{+0.7}$ & $2.0_{-0.4}^{+0.3}$ & $0.8_{-0.4}^{+0.5}$\\
59079 & 3033290113,114 & $0.9\pm0.1$ & $1.1_{-0.3}^{+0.6}$ & $2.4_{-0.4}^{+0.3}$ & $1.4_{-0.4}^{+0.5}$\\
59081 & 3033290115,116 & $1.06_{-0.11}^{+0.09}$ & $0.7\pm0.2$ & $3.0_{-0.3}^{+0.4}$ & $2.1\pm0.6$\\
59086 & 3598010901 & $0.96_{-0.10}^{+0.08}$ & $0.8_{-0.2}^{+0.3}$ & $2.0_{-0.2}^{+0.3}$ & $1.6_{-0.4}^{+0.3}$\\
59088 & 3033290117,118 & $0.8\pm0.1$ & $1.3_{-0.5}^{+0.9}$ & $1.9_{-0.4}^{+0.3}$ & $1.1\pm0.2$\\
59092 & 3033290119,120 & $0.7\pm0.2$ & $2_{-1}^{+4}$ & $1.7_{-0.6}^{+0.3}$ & $0.7\pm0.4$\\
59098 & 3033290122,123 & $0.5\pm0.2$ & $3_{-2}^{+14}$ & $2.2_{-0.5}^{+0.2}$ & $0.5_{-0.4}^{+0.3}$\\
59102 & 3033290124,125,126 & $0.5_{-0.1}^{+0.2}$ & $4_{-2}^{+9}$ & $1.7_{-0.5}^{+0.2}$ & $0.6_{-0.4}^{+0.3}$\\
59122 & 3033290128,131,132 & $0.8_{-0.2}^{+0.3}$ & $0.7_{-0.4}^{+2.0}$ & $1.3_{-0.5}^{+0.3}$ & $0.5_{-0.3}^{+0.4}$\\
59137 & 3033290135,136,137 & $0.8_{-0.1}^{+0.2}$ & $0.8_{-0.4}^{+0.9}$ & $1.4_{-0.4}^{+0.3}$ & $0.6\pm0.3$\\
59145 & 3598011101 & $0.5_{-0.2}^{+0.5}$ & $<28$ & $1.2\pm0.2$ & $<0.4$\\
59154 & 3033290139,141,143 & $0.3_{-0.1}^{+0.2}$ & $<6000$ & $1.5\pm0.4$ & $<0.8$\\
59165 & 3033290114,146,147 & $0.14_{-0.04}^{+0.12}$ & $<1\times10^8$ & $1.8_{-0.8}^{+1.4}$ & $<2.2$\\
59174 & 3033290150,3598011201,202 & $0.3_{-0.2}^{+0.4}$ & $<3\times10^7$ & $0.8\pm0.2$ & $<1.1$
\enddata
\tablecomments{The fluxes are calculated in the energy range of 1--10 keV.}
\end{deluxetable*}

\section{Discussion}  \label{sec:diss}

The August 2020 X-ray outburst was the second magnetar-like outburst from PSR \psr since its discovery in 1999, after a 14 year hiatus since its first recorded outburst in 2006.  Our \nicer\ GO and ToO observations provided continuous monitoring and timing of the source starting about three and a half years before the 2020 outburst until more than a year following the outburst in November 2021.  Our monitoring observations prior to the August 1, 2020 \swift\ announcement of the outburst showed that the flux was already increasing above quiescent level by late June.  Due to our high cadence observations during and following the outburst, we were able to resolve rapid changes in the pulsed flux, which our measurements showed increased by more than a factor of ten above the quiescent level.  While this flux increase is much higher than the five-times increase reported for the 2006 outburst \citep{Kuiper2009}, the flux during that outburst was less densely sampled and timing coherence was lost \citep{Gavriil2008}, so higher flux peaks may have been missed.  Both outbursts were accompanied by spin-up glitches of very similar size and in both cases, transient thermal components appeared that were responsible for most of the flux increases.  The thermal components and the flux also decayed over similar timescales of several months.  However, we observed a change in the X-ray pulse profile during the 2020 outburst, while no change in profile was reported for the 2006 outburst.  Overall, the characteristics of the two magnetar-like outbursts of \psr are very similar.

PSR \psr thus appears to be a unique and fascinating source with a dual nature, behaving most of the time like a RPP and occasionally like a magnetar.  It is also an unusual young RPP, having no detected radio or GeV emission.   So is it a RPP that is becoming a magnetar, or a magnetar masquerading as a RPP?  As we discuss below, the quiescent behavior as well as the outbursts can be interpreted in either scenario but involve very different physical mechanisms for the emission.

First, we consider a pulsar-like model for PSR \psr.  \cite{Harding2017} presented a model for MeV pulsars like \psr in which their lack of radio or GeV emission is due to geometry.  In this model, MeV pulsars are the same as the rest of the young RPP population: they have global dipole magnetospheres that are nearly force-free, emit radio emission from above the magnetic poles and $\gamma$-ray and non-thermal X-ray emission from near the current sheet outside the light cylinder \citep{Kala2018,Harding2021}.   However, they have small magnetic inclination angles and we view them along lines-of-sight, away from the magnetic poles, that miss both the radio beams and the center of the current sheet where the highest energy acceleration occurs and where the GeV emission is produced.  In this model then, the quiescent X-ray emission is synchrotron radiation (SR) from electron-positron pairs near the current sheet, where the shape of the SED derives from the smoothly falling spectrum of the pairs \citep{Harding2015,Harding2021}.  The magnetar-like outbursts of \psr are due to the appearance of or a change in a toroidal surface magnetic field component that provides a field twist, and more particle acceleration and pair production near the surface.  The increased pair multiplicity increases the polar cap heating and opens up more field lines, increasing the size of the polar caps.  If our line-of-sight then crosses nearer to the larger heated polar cap, we will temporarily see a thermal component, but we can still miss the radio emission since it is more tightly beamed than the thermal emission.  However, in this case the thermal component will not be in phase with the non-thermal SR pulse which is a major drawback for this model.  The measurements of a braking index in \psr \citep{Livingstone2011} with values typical of RPPs may add support to the RPP model. However, it is unclear what we should expect for  magnetar braking indices and why they have not been detected. 

Next, we consider a magnetar-like model in which the quiescent X-ray emission from \psr is resonant inverse-Compton scattering emission (RICS) from particles accelerated by field twists and scattering thermal photons from the hot neutron star surface \citep{Baring2007,Belo2013}.  We do not see any GeV emission because photon splitting and pair production attenuate the RICS spectrum above about 1 MeV \citep{Wadiasingh2019}.  The outbursts are then typical magnetar bursts where crustal motions cause increases in the magnetic field twists, particle acceleration and RICS emission.  The increased particle acceleration and pair/splitting cascades produce visible heated spots due to particle surface bombardment \citep{Belo2013}.  Since this emission occurs on closed field lines near the equator, we could miss any transient radio emission along open field lines.  There are several potential drawbacks in this model.  If the quiescent emission is RICS, why don't we see a quiescent thermal component as appears in magnetars?  Supporting the non-observed thermal component, many/most young magnetars have X-ray luminosities $10^{34}-10^{35}\mbox{ erg s$^{-1}$}$ (see eg. magnetar review by \citet{Kaspi2017}), which would translate to fluxes $\sim 2\times 10^{-12}-10^{-11}\,\rm erg\,s^{-1}\,cm^{-2}$ at 6.5 kpc, lower than the observed fluxes in Table 3. Perhaps this component is much weaker or non-existent for \psr because its magnetic activity is not persistent enough to produce heated spots from particle bombardment, and with the geometry for \psr proposed above, we would not see the heated polar caps in quiescence.  In addition,  there is no significant change in flux or spectrum of the power-law component during the outburst.  Also, magnetar pulse profiles are typically more complex than that of \psr\ \citep{HuNH2019}  but this could result from less persistent magnetic field activity as well.  In addition, the X-ray luminosity of \psr is less than its spin down power, whereas for magnetars the X-ray luminosity exceeds the spin-down power by several orders of magnitude.  Finally, one could argue that the SED of \psr is closer to the smooth SR-like spectrum of RPPs like the Crab and B1509$-$58 \citep{Kuiper2018} that do not show the sharp cutoff predicted for magnetar SEDs \citep{Wadiasingh2019}.  However, at the moment there are no detected spectral points in the SED of \psr between 200 keV and 30 MeV so a sharp attenuation cutoff could exist.   In addition, both the quiescent and outburst X-ray spectra of \psr is very close to the spectrum of magnetar RXS J1708$-$40 \citep{Kuiper2009}.  The true test then of a pulsar-like vs. a magnetar-like model for PSR \psr is a measurement of its SED between 200 keV and 30 MeV.

Regardless of whether the pulsar-like or magnetar-like model for \psr is more plausible, the existence and behavior of \psr and J1119$-$6127 suggest a continuum of neutron stars that lie between RPPs and magnetars.   The enhanced quiescent thermal emission observed from them and several other high-B RPPs supports such a continuum \citep{Kaspi2005, NgKH2012, Olausen2013, HuNT2017}. In fact, the characteristics of \psr are similar to those of transient Anomalous X-Ray Pulsars \citep{Livingstone2010} and may indicate a link between RPPs and magnetars.  An additional similarity between RPPs and magnetars is the discovery of a wind nebula around the canonical magnetar Swift J1834.9$-$0846 \citep{Younes2016}.  What is less clear is their direction of evolution.  \citet{Espinoza2011,Espinoza2022} showed that the high-B pulsar J1734$-$333 is evolving towards the magnetars (based on its braking index). There is no evidence for magnetar X-ray behavior for this pulsar, but this result may indicate the direction of evolution.
Since both \psr and J1119$-$6127 are very young, it is possible that a stronger interior magnetic field is in the process of emerging over a timescale $< 10^3 - 10^4$ yr \citep{Bhattacharya2008}.   After the 2006 outburst of \psr, its  braking index was observed to decrease by 18\%  \citep{Livingstone2011}, which could be caused by an increase in magnetic field strength \citep{Blandford1983, Ho2015}. Yet, such sources may have been born with stronger magnetic fields that are in the final stage of decaying \citep{PonsMG2009, ViganoRP2013, GourgouliatosWH2016}.  
Continued X-ray monitoring of \psr as well as expanded spectral coverage will be crucial in understanding of its true nature and place in neutron star evolution.

\begin{acknowledgments}
This work was supported by the National Aeronautics and Space Administration (NASA) through the \nicer\ mission, the Astrophysics Explorers Program and through the \nicer\ Guest Investigator Program. 
This work partly made use of data supplied by the UK \swift\ Science Data Centre at the University of Leicester.
This research has also made use of data obtained with \emph{NuSTAR}, a project led by Caltech, funded by NASA and managed by NASA/JPL, and has utilized the NUSTARDAS software package, jointly developed by the ASDC (Italy) and Caltech (USA).
C.-P.H.~acknowledges support from the National Science and Technology Council in Taiwan through grant MOST 109-2112-M-018-009-MY3. 
W.C.G.H.~acknowledges support through grants 80NSSC21K2035 and 80NSSC22K1308 from NASA.
C.M.E.~acknowledges support from the grant ANID FONDECYT 1211964.

\end{acknowledgments}

\facilities{\emph{NICER}, \emph{NuSTAR}, \emph{Swift} (XRT)}
\software{HEASoft, XSPEC}

\bibliography{J1846}{}
\bibliographystyle{aasjournal}

\end{document}